\documentstyle[prl,aps,multicol]{revtex} \tighten 
\begin{document} \draft
\title{$1/f$ noise in variable 
range hopping conduction.}
\author{ B. I. Shklovskii} 
\address{Theoretical Physics
Institute, University of Minnesota, 
116 Church St. Southeast, Minneapolis,
Minnesota 55455}
\maketitle
\begin{abstract}
$1/f$ noise induced by traps consisting of donors 
with no neighbors with close energies in their vicinity is studied.
Such donors slowly exchange electrons with the rest of 
conducting media. It is shown that in the 
variable range hopping regime $1/f$ 
noise exponentially grows with the decreasing temperature. 
At high temperatures, when the variable range hopping 
crosses over to the nearest neighbor one, 
we predict a very weak temperature dependence 
in spite of the activation dependence of the conductivity. 
\end{abstract}

\section{Introduction}

\begin{multicols}{2}

Low frequency $1/f$ noise was found in 
many conducting materials with the wide 
variety of transport mechanisms.
Substantial attention was devoted to 
$1/f$ noise in the hopping 
conduction\cite{Voss,Cohen,Shlimak,Lee,McCammon,Savchenko,Shk,KoShk,Kozub,Yu,Kogan}.
One of the reasons is that $1/f$ noise limits performance of 
important devices working in the hopping regime, for example, 
thermistors for X-ray detection and 
other astrophysical applications\cite{McCammon}.
Understanding of the nature of $1/f$ noise can also shed 
light on the role of electron-electron interactions
in the hopping transport.

In a lightly doped 
semiconductor at low temperatures 
electrons are localized on
donors and the conductivity 
is due to hopping between donors.
(For concreteness we talk about
$n$-type moderately compensated 
semiconductor with concentration of donors $N_D$.)
There are two well known
regimes of hopping transport.
At relatively high temperatures electrons 
use almost all the donors for hopping. This regime is called 
the nearest neighbor hopping (NNH). 
In this case, the energy scatter of donors 
due to random potentials of charged 
donors and acceptors plays only a secondary role, 
leading to a fixed activation energy of conductivity.
At low enough temperatures activation required for the 
use of the majority of donors becomes 
very costly and only donors within a narrow band of energies 
around the Fermi level participate in the conductivity.
With decreasing temperature the width of this band 
shrinks and hops become longer. This explains the name 
"variable range hopping" (VRH) of the regime.

Most of recent experiments deal 
with VRH and 
focus on temperature dependence of $1/f$ noise
\cite{Cohen,Lee,McCammon,Savchenko}.
Results seem to be quite controversial. 
For example, Lee\cite{Lee} found that in silicon 
$1/f$ noise amplitude decreases when temperature 
goes down,
while McCammon\cite{McCammon} observed 
exponential increase of $1/f$ noise 
with the decreasing temperature.

On the other hand, the original theory of $1/f$ noise
in the hopping transport\cite{Shk,KoShk} 
deals only with NNH and only with the case of
 relatively high temperatures
when NNH conductivity is temperature independent.
This theory is based on the 
idea of traps provided by rare isolated 
donors. Such a donor traps  
an electron from donors of the transport
paths (conducting media) and releases it back, both 
with characteristic times, which are much larger 
than times of hops determining VRH conductivity.
The resulting modulation 
of number of "conducting" carriers 
at low frequency $\omega$ 
leads to the spectral
density of current noise $I^{2}_{\omega}$
with a behavior close to $1/\omega$ (or $1/f$).
This idea is similar to the McWorter's explanation 
of $1/f$ noise in MOSFETs by electron 
exchange between two-dimensional electron gas and traps in the oxide,
which have an extremely wide spectrum 
of times of tunneling to the interface\cite{McW,Koganbook}. 
The difference is that in the 
MOSFET case traps are separated 
from the conducting two-dimensional gas by the
interface, while in the case of the hopping transport
traps are located inside pores scattered in the body 
of the conducting media of donors.
This paper generalizes the idea of Refs.\onlinecite{Shk,KoShk} to the 
VRH conductivity
(Sec. II) and to the temperature dependent NNH (Sec. III). 
 We again identify those donors,
which can work as traps with extremely wide spectrum of very large
relaxation times. Then we calculate the probability of a trap,
which now becomes an exponentially decreasing function of 
temperature. 
Integrating contributions of all traps to the noise
we arrive at the conclusion that in the VRH case 
the spectral density of current noise $I^{2}_{\omega}$ approximately
obeys the Hooge law, Eq.~(\ref{Hooge}),
with the coefficient $\alpha$, which is given by Eq.~(\ref{alpha}) and
exponentially decreases with temperature.
On the other hand, in the case of NNH conductivity, 
which grows exponentially with temperature 
with a constant activation energy, the Hooge's $\alpha$ 
is almost temperature independent.
Thus, exponential 
temperature dependence of conductivity 
does not automatically lead to 
exponential dependence of $1/f$ noise.
 
Sec. IV is devoted to the discussion 
of the relationship between
of spectral densities of fluctuations of 
the current and the electron concentration,
which is used in Sec.II and Sec. III 
to calculate the noise spectral density.
In Sec. V. we compare our theory 
with two previously published theoretical 
works\cite{Kozub,Yu} on $1/f$ noise in VRH
and with available experiments\cite{Cohen,Lee,McCammon,Savchenko}.
We also make a comment
about applicability of our theory to MOSFETs.

\section{Variable range hopping}

Due to the Coulomb interaction of localized electrons,
the density of states of donor states, $g(\varepsilon)$,
of a three-dimensional lightly doped semiconductor 
is known\cite{ES75,SE} 
to have the Coulomb gap near the Fermi level, 
\begin{equation}
g(\varepsilon) = (3/\pi) \kappa^{3} e^{-6} \varepsilon^{2}~.
\label{g}
\end{equation}
Here $\varepsilon$ is the energy of 
a localized electron on a donor
calculated from the Fermi level, 
$\kappa$ is the dielectric constant of the semiconductor and
$e$ is the proton charge.
At low temperatures VRH conductivity $\sigma(T)$ 
obeys the Efros-Shklovskii (ES) law\cite{ES75,SE} 
\begin{equation}
\sigma(T) = \sigma_0\exp[-(T_0/T)^{1/2}],
\label{sigma}
\end{equation}
where $T_0 =Ce^2/k_B\kappa a$, $a$ is the localization length 
(Bohr radius) 
of an electron on a donor, and $C \simeq 2.7$. 

The ES conductivity 
is determined only by donors with energies in the band 
$\delta = k_B (T_{0} T)^{1/2}$ around the Fermi level ($\delta$-band). 
Similarly to Ref.\onlinecite{Shk,KoShk} 
we are interested in a trap, i. e. in such a rare donor of the same band 
$\delta$, which anomalously slowly exchanges an electron with the 
majority of donors of this conducting band. 
The rate of electron hops 
between two donors $i$ and $j$ is\cite{SE} 
\begin{equation}
\nu_{ij} = \nu_{0}\exp\left(-\frac{2r_{ij}}{a}
 - \frac{\varepsilon_{ij}}{k_BT}\right), 
\label{rate}
\end{equation}
where $\varepsilon_{i}$ is the energy level of the 
donor $i$, $\varepsilon_{ij} = (|\varepsilon_{i}| + |\varepsilon_{j}| +
|\varepsilon_{i} - \varepsilon_{j}|)/2$.
We want to find a trap with relaxation time larger than $\nu^{-1}$,
where $\nu$ is so small that $\ln(\nu_{0}/\nu)\gg (T_0/T)^{1/2}$.
Such a trap is an isolated donor, for which all rates of 
transitions to the neighboring donors of the band of energies $\delta$
are smaller than $\nu$.
Let us consider a donor $i$ located 
in the center of a sphere of radius
$R(\nu) =(a/2)\ln(\nu_{0}/\nu)$, which contains no other
donors $j$ within $\delta$-band.
Then there are no direct hops from our donor 
to the conducting $\delta$-band of donors.
But this is not enough to have a trap we are looking for, 
because an electron can still escape from donor $i$ to the 
conducting $\delta$-band through 
an intermediate donor $j$, which has energy $\varepsilon_{j}$ 
much larger than $\delta$,
but is located closer  to $i$, at the distance $r_{ij} < R(\nu)$.
In order to preserve the trap all such "dangerous"
intermediate donors $j$ near donor $i$ 
should be eliminated or, in other words, all the remaining ones 
should satisfy inequality
\begin{equation}
\frac{\varepsilon_{ij}}{k_BT} 
> \ln\frac{\nu_{0}}{\nu}.
\label{pore}
\end{equation}
This inequality means that 
the energy $\varepsilon_j$ should be away from
the energy band of the width
$\Delta = k_BT\ln(\nu_{0}/\nu) \gg \delta$ 
around the Fermi level ($\Delta$-band).
The concentration of 
donors in $\Delta$-band is of the order of 
$N(\Delta) \sim g(\Delta)\Delta= (\kappa \Delta/ e^2)^{3}$.
Therefore, the average number of $\Delta$-band donors
in the sphere of 
radius $R(\nu)$ equals $M =(4\pi/3)R^{3}(\nu)N(\Delta)$.
Probability, $W(\nu)$,
to find a sphere with radius $R(\nu)$ empty from  
such donors has a form $\exp(-M)$ or
\begin{equation}
W(\nu) = \exp\left(- B(T/T_0)^{3}
\ln^{6}\frac{\nu_{0}}{\nu}\right)~,
\label{probability}
\end{equation}
where $B$ is a numerical factor.
For a donor with the relaxation rate $\nu$ 
equal the typical hopping frequency 
of conducting electrons,
$\nu_c \sim \nu_0 \exp[-(T_0/T)^{1/2}]$,
we get $W(\nu) \sim 1$, confirming that 
this is a typical donor.  
Traps have $\nu \ll \nu_c$ and, therefore,
exponentially small probabilities.
To find the coefficient $B$ we 
calculated probability of a pore
in four dimensional space of three coordinates and energy.
This calculation gives $B \simeq 0.3$.
Note that probability $W(\nu)$ decreases 
with increasing temperature because at high
temperatures the volume of the four 
dimensional pore 
from which one has to eliminate donors becomes larger. 
 
The spectral density of fluctuations of the concentration 
of "conducting" electrons, $n^{2}_{\omega}$,
can be evaluated by integration over all traps\cite{Koganbook} 
\begin{equation}
n^{2}_{\omega} =\frac{N(\delta)}{V}
\int_0^{\nu_c}\frac{dW}{d\nu}\frac{4\nu}{\omega^{2} + \nu^{2}}d\nu.
\label{n}
\end{equation}
Here $N(\delta)$ is the 
concentration of donors in the $\delta$-band
around the Fermi level and $V$ is the sample volume. 

According to the standard approximation 
is that relative fluctuations of the 
concentration of conducting electrons 
lead to 
fluctuations of conductivity, i.e.
\begin{equation}
\frac{I^{2}_{\omega}}{I^{2}} \propto
\frac{n^{2}_{\omega}}{N^{2}(\delta)}~.
\label{nI2}
\end{equation}
Here $I^{2}_{\omega}$ is the spectral 
density of current fluctuations and $I$ is the average current 
through the sample. We will discuss justification 
of Eq.~(\ref{nI2}) in Sec. IV.
Using Eq.~(\ref{n}) we can now 
estimate $I^{2}_{\omega}/I^{2}$. 
Let us start from the most interesting case when 
the exponential factor $dW/d\nu$
in the integrand of Eq.~(\ref{n}) 
is a weaker function 
of $\nu$, than $\nu/(\omega^{2} + \nu^{2})$
and the integral is determined by $\nu \sim \omega$.
Then one can take exponential term of Eq.~(\ref{n}) out of the 
integral at the frequency $\nu = \omega$. 
This leads to the Hooge law\cite{Hooge}
\begin{equation}
\frac{I^{2}_{\omega}}{I^{2}} = \frac{\alpha(\omega, T)}{\omega N_{D}V},
\label{Hooge}
\end{equation}
where $N_{D}V$ is the total number of donors and 
$\alpha$ is the Hooge's coefficient. We obtain 
\begin{equation}
\alpha(\omega, T) \propto
\exp[- B(T/T_0)^{3}\ln^{6}\frac{\nu_{0}}{\omega}].
\label{alpha}
\end{equation}
Here as everywhere below we 
are concentrated on exponential temperature dependence of 
noise and are not trying to calculate prefactor. 
Using Eq.~(\ref{sigma}) one can express $\alpha(\omega, T)$ 
through $\sigma(T)$:
\begin{equation}
\ln\alpha(\omega, T) = - B
\left(\frac{\ln(\nu_{0}/\omega)}{\ln(\sigma_0/\sigma)}\right)^{6}.
\label{alphasigma}
\end{equation}
This simple relationship can be convenient for a 
direct experimental verification.

The above derivation of $\alpha(\omega, T)$ 
is justified when the dependence of
$\alpha$ on $\omega$ is weaker than $1/\omega$,
i. e. when $\omega \gg  \nu_0 \exp[-(T_0/T)^{3/5}]$.
On the other hand, our 
approach of traps is good only for 
frequencies $\omega$ smaller than the 
frequency of transport hops $\nu_c$, i. e.
for  $\ln(\nu_{0}/\omega) \gg (T_0/T)^{1/2}$. Thus, 
the Hooge law is valid in the range of frequencies
\begin{equation}
\exp[-(T_0/T)^{1/2}] \gg  \omega/\nu_0 \gg  \exp[-(T_0/T)^{3/5}]~.
\label{range}
\end{equation}
In spite of close powers in the exponents of lower 
and upper limit this range can be quite broad.
For example, even at a moderate $(T_0/T)^{1/2}= 10$ 
this range covers 2.5 decades, while at the very 
large $(T_0/T)^{1/2}= 20$ it reaches 7 decades.

One can approximately write
$I^{2}_{\omega}/I^2 \propto 1/\omega^{\gamma}$,
where $\gamma = 1 - 6B(T/T_0)^{3}\ln^{5}(\nu_{0}/\omega)$.
Thus, according to this theory $\gamma < 1$. In the range 
(\ref{range}), the index $\gamma$ is close to unity and approaches 
1/2 at the low frequency limit of applicability of 
Eq.~(\ref{Hooge}). 
 
For a fixed $\omega$ in the frequency range ((\ref{range})),
Eq.~(\ref{alpha})
contains the main result of this paper - an exponential
growth of the noise power with the decreasing temperature.
The physics of this dependence is clear: the lower 
the temperature, the smaller is the width 
of the band of energies $\Delta$, where 
one has to eliminate  'dangerous" 
donors in order to get a trap for a given frequency,
the smaller is the number of donors to be eliminated,
the larger is the probability to find such a trap.

According to Eqs.~(\ref{nI2}) and (\ref{n}) at
smaller frequencies $\omega \ll \nu_{min} = 
\nu_0 \exp[-(T_0/T)^{3/5}]$, the
spectral density of noise becomes substantially 
weaker than $1/f$ and eventually saturates 
at the level $\sim  \exp[(T_0/T)^{3/5}]$.
To calculate behavior of $1/f$ noise in all range 
$\omega < \nu_c$ more accurately one should similarly to\cite{Shk,KoShk}
add the contribution of traps made of large finite 
clusters of donors
and calculate the integral of Eq.~(\ref{n}) numerically.

We derived Eq.~(\ref{alpha})
for a semiconductor in which 
the concentration $N_{D}$ is substantially smaller
than the critical concentration of the metal-insulator
transition, $N_{MI}$. In this case, we can use
Eq.~(\ref{g}) for the density of states $g(\varepsilon)$.
In experiments VRH is typically measured 
at $0.25 N_{MI} \leq N_{D} \leq N_{MI}$
because of a huge resistance of lighter doped samples.
We believe that Eqs.~(\ref{Hooge}) and (\ref{alpha}) 
still work in the range 
$0.25 N_{MI} \leq N_{D} \leq 0.5 N_{MI}$.

We do not exactly know how to generalize Eq.~(\ref{alpha}) 
for the critical 
vicinity of the metal-insulator transition, 
where  $N_{MI}- N_{D} \ll N_{D}$.
An additional source of the traps may be
related to the spacial fluctuations  of 
diverging at the transition localization length 
$\xi$. Anomalously small $\xi$ in a vicinity of some 
donor can create a trap.
A simple conjecture is that Eqs.~(\ref{alpha}) and (\ref{alphasigma}) 
remain valid at
$N_{MI}- N_{D} \ll N_{D}$  with 
the same $T_0 =2.7 e^2/k_{B}\kappa \xi$ as 
in the ES law near the transition. Anyway an experimental 
study of this range of concentration can shed some light 
on the metal-insulator transition.

Until now we have been dealing with VRH conductivity 
in a doped crystalline moderately compensated semiconductor, 
where conductivity obeys Eq.~(\ref{sigma}).
In systems with a weaker role of the electron-electron 
interaction such 
as amorphous semiconductors or two-dimensional systems screened
by a parallel metallic gate, VRH can take place in the larger 
band of energies than the width of the 
Coulomb gap and, therefore, one can observe Mott law
\begin{equation}
\sigma(T) = \sigma_0\exp[-(T_{d}/T)^{1/(d+1)}],
\label{sigma}
\end{equation}
where $d= 2, 3$ is the dimensionality of 
the system, $T_{d}= \beta_{d}/(ga^{d})$,
$g$ is the bare 
(unperturbed by the Coulomb interaction) 
$d$-dimensional density
of states in the vicinity of the Fermi level 
and $\beta_{d}$ are numerical constants\cite{SE}.
Repeating the derivation of Eq.~(\ref{alpha}) 
we get the Hooge law with 
\begin{equation}
\alpha(\omega, T) \sim
\exp[- B_{d}(T/T_d)\ln^{d+1}(\nu_{0}/\omega)]~,
\label{alpha2}
\end{equation}
or 
\begin{equation}
\ln\alpha(\omega, T) = - B_d
\left(\frac{\ln(\nu_{0}/\omega)}
{\ln(\sigma_0/\sigma)}\right)^{d+1},
\label{alphasigma1}
\end{equation}
where $B_d$ are numerical coefficients.
The physics of the origin of the exponential temperature
dependence is exactly the same as above:
the higher the temperature, the more difficult is
to find a trap for a given small frequency $\nu$.

\section{Temperature dependent nearest neighbor hopping}

Eqs.~(\ref{alpha}) is valid when 
temperature is so small that the energy width $\Delta$   
of the band, where all "dangerous" donors are eliminated,
is smaller than the width of the donor band, $A=e^{2}N^{1/3}_{D}/\kappa$, 
i. e. at $k_BT\ln(\nu_{0}/\nu) \ll e^{2}N^{1/3}_{D}/\kappa$.
In the opposite case, $\Delta \gg A$, or
\begin{equation}
T \gg T_1= \frac{e^{2}N^{1/3}_{D}}{k_B\kappa\ln(\nu_{0}/\nu)},
\label{T1}
\end{equation}
the concentration of donors in the $\Delta$-band 
\begin{equation}
N(\Delta) = \int_{-\Delta}^{\Delta}g(\varepsilon)d\varepsilon =
N_{D}[1-(A/\Delta)^3]
\label{NDelta}
\end{equation}
is close to $N_{D}$ 
and only weakly depends on $\Delta$.
Here we used the fact that as shown in 
Chapter 3 of Ref.\onlinecite{SE} the large energy tails of 
the density of states of the classical donor band 
behave as
$g(\varepsilon)\sim N_D A^{3}/|\varepsilon|^{4}$ 
at $|\varepsilon|\gg A$. 
Now we can use Eq.~(\ref{n}) with probability 
$W(\nu)=\exp[-(\pi/6) N(\Delta)a^3\ln^{3}(\nu_{0}/\nu)]$ 
which can be rewritten as 
\begin{equation}
W(\nu) = \exp[-(\pi/6) N_{D}a^3\ln^{3}(\nu_{0}/\nu)]\exp(T_c/T)^{3}~.
\label{newW}
\end{equation}
Here $T_{c} \simeq e^{2} a N^{2/3}_{D}/k_B \kappa$ 
is the critical temperature of 
the transition from NNH to VRH\cite{SE}. 
In the range of frequencies
\begin{equation}
\exp[-(T_0/T)^{1/2}] \gg  \omega/\nu_0 \gg  \exp[-(N_{D}a^3)^{-1/2}]~,
\label{rangef}
\end{equation}
where $dW/d\nu$ is weaker function of $\nu$ than
 $4\nu/(\omega^{2} + \nu^{2})$.
we again obtain the Hooge law with 
\begin{equation}
\alpha(\omega,T) =
\exp[-(\pi/6)N_{D}a^{3}\ln^{3}(\nu_{0}/\omega)] \exp(T_c/T)^{3}~.
\label{alpha1}
\end{equation}
Eq.~(\ref{alpha1}) is valid both
for NNH (at $T > T_{c}$) and for VRH (at $T_1 < T < T_{c}$).
First exponential factor of Eq.~(\ref{alpha1}) 
originates from the probability of a pore free of all donors.
This factor is temperature 
independent and coincides with the value of $\alpha$ 
obtained for the high temperature limit of NNH~\cite{Shk,KoShk}.
Second, temperature dependent exponential factor 
of the right hand side of Eq.~(\ref{alpha1})
provides only a relatively small correction 
to $\ln\alpha(\omega,T)$, which grows
as $1/T^3$ with the decreasing temperature. 
In NNH range, when $T \gg T_{c}$, 
the relative temperature dependent correction
to $\alpha(\omega,T)$ is much smaller than unity.
At the transition point from NNH to VRH, i. e. at $T \sim T_{c}$, 
this correction reaches 100\%. At $T \ll T_{c}$
it becomes exponentially large and
provides the smooth crossover to Eq.~(\ref{alpha}), 
when $T$ approaches
$T_1=e^{2}N^{1/3}_{D}/k_B\kappa \ln(\nu_{0}/\nu)$.
However, as we already mentioned above, 
because of huge resistances an experimental observation of VRH 
is not possible, when $N_{D}a^{3}\ll 1 $.
Therefore, inequality (\ref{rangef}) is very difficult to realize
and the use of Eq.~(\ref{alpha1}) for VRH regime is 
limited. But Eq.~(\ref{alpha1}) is definitely important 
for the temperature dependence of noise of NNH,
which takes place at $N_{D}a^{3}\ll 1$. 

To conclude this section
we would like to repeat that for NNH temperature 
dependence of $\alpha(\omega,T)$ is negligible
in all range of temperatures 
$A\gg T \gg T_c$,
where conductivity itself grows with decreasing 
temperature with activation energy of the order of $A$\cite{SE}.

\section{Relationship between fluctuations
of the current and of the concentration of conducting electrons}

Now we would like to return to Eq.~(\ref{nI2}), 
which establishes relation between 
fluctuations of concentration of conducting 
electrons (which are not trapped) $n$
and fluctuations of current 
and discuss its validity.
In MOSFETs, where most of electrons are free, 
fluctuations of the concentration
of free electrons obviously lead to proportional 
fluctuations of the conductivity\cite{McW,Koganbook}
and at a fixed electric field $I^{2}_{\omega}/I^{2} =
n^{2}_{\omega}/n^{2}$.

For NNH such proportionality between 
fluctuations of $I$ and $n$ is obvious for strongly
compensated samples, where electrons play the role of carriers, or for a
weakly compensated semiconductor,
where the role of carriers is played by a small concentration of 
holes (empty donors).
However, even in this case
there could be such an intermediate compensation
ratio $K= N_{A}/N_{D}$,
where the coefficient in linear proportionality between 
fluctuations of $I$ and $n$  vanishes. 
(Here $N_{A}$ is the concentration of acceptors). 
In VRH conductivity connection between
fluctuations of $I$ and $n$ is even less trivial.

Let us consider relation of $I$ and $n$ for
a simpler example when 
concentration, $n$, of electrons in a two-dimensional hopping system 
is varied due to a change of the gate voltage.
Any variation of $n$ leads to a shift of 
the Fermi level. The Coulomb gap, however, 
moves together with the Fermi level and, in the first 
approximation, it remains unchanged at small energies $\delta$,
which are important for VRH. It is not obvious then
whether Eq.~(\ref{nI2}) is valid.

In the first subsection of this section 
we show that in the generic case 
of asymmetric with respect of Fermi
level bare (disorder related) density 
of states one can justify
Eq.~(\ref{nI2}), because in this case
there is a nonzero derivative $d\sigma/dn$ for 
a sample of infinite size. In the second subsection 
we introduce more general mechanism of fluctuation 
of conductivity, which exists even in the case
when $d\sigma/dn=0$ for infinite sample, 
but still can be evaluated in the way 
it was done in Secs. II and III.

\subsection{Asymmetry of the density of states}

In a doped $n$-type semiconductor at a generic 
compensation ratio $K$
the density of states at large $\varepsilon$
 is asymmetric with respect to the Fermi level
(see plots in Chapter 14 of Ref. \onlinecite{SE}).
In spite of tendency of Coulomb gap
to be symmetric,
the density of states, $g(\varepsilon)$,
at $\varepsilon \sim \delta$, is sensitive to
behavior of the density of states
at larger energies.
This happens because 
the Coulomb gap should
crossover in its "shoulders" to the bare density of 
states of the classical impurity band\cite{SE}.
Using the self-consistent equation 
for the one-electron 
density of states derived by Efros\cite{Efros} 
one can show that  
the absolute value of nonuniversal
(depending on position of Fermi level)
 correction to the density states
at small $\varepsilon$ is proportional to $\kappa^{3}\varepsilon^{3}/e^{6}A$,
where $A \simeq e^{2}N^{1/3}_{D}/\kappa$
is the energy width of the classical donor band.
In other words, the relative asymmetric correction to 
Eq.~(\ref{g}) at $\varepsilon \sim \delta$ is proportional 
to $\delta/A$.

Our final goal is to evaluate fluctuations of the hopping conductivity. 
In VRH regime the conductivity fluctuates because $g(\delta)$ 
fluctuates following the 
Fermi level. It is easy to show that 
relative fluctuations of the conductivity are larger than fluctuations
of $g(\delta)$  by $(T_0/T)^{1/2} \sim T_0/\delta$.
Thus, as in a free electron 
gas, the fluctuations of conductivity are 
proportional to fluctuations of the concentration of conducting electrons
with the temperature independent coefficient of proportionality 
$(\delta/A)(T_0/\delta) = T_0/A$. 

Now we can return to slow traps immersed 
in the conducting media, which
work similarly to a gate. While the number of trapped electrons 
is slowly changing with time,
the conducting media arrives at the
quasi-equilibrium at much shorter time scale 
and develops the quasi-equilibrium Fermi level tracking  
fluctuations of the electron concentration. In turn, the 
position of the quasi-Fermi level
affects the density of states 
relevant to the conductivity and, therefore, 
the hopping conductivity itself.

Without Coulomb interactions 
fluctuations of the concentration 
are uniform over the conducting media.
Coulomb interactions makes these
fluctuations localized, because 
of screening of trapped electrons. 
When an electron is trapped, a hole remaining
in the conducting media is localized 
close to the trap. In other words,  
fluctuations of the concentration of conducting electrons do not propagate 
to the bulk of the conducting media. In this sense, traps
are equivalent to small gates 
separated from each other by large distances.
They locally perturb the concentration of conducting electrons.
This perturbation, of course, changes conductivity 
only near traps. Now we have to discuss these changes
and the way how they affect the total current through the sample.

We start from an estimate of the screening radius 
of a spherical trap with radius 
$R = (a/2)\ln(\nu_0/\nu)$. 
In other words, we want to find the radius 
of the concentric sphere, where the hole left 
behind by trapped electron is localized.
In the Coulomb gap, the density of states 
at the Fermi level is very small at small temperatures.
One can find substituting $k_BT$ for $\varepsilon$
into Eq.~(\ref{g})
so that linear screening Debye radius $R_{s}$ which describes 
screening of very small charge is very large, $R_{s} \sim a(T_0/T) $.
On the other hand, one electron charge $-e$ of the 
trap is large enough to produce additional
nonlinear screening. Nonlinear screening 
radius $R_{N}$ can be found from the condition that 
potential of trapped electron $-e/ \kappa R$ 
can attract a positive charge to one of donors inside 
the sphere with radius $R_{N}$. This condition can be written as 
\begin{equation}
\frac{4\pi}{3}(r_{N}^{3}-R^{3})
\int_{0}^{e/\kappa R}2g(\varepsilon)d\varepsilon \sim 1~,
\label{R_{N}}
\end{equation}
This gives $R_{N}- R \sim R$. 
To understand changes of the macroscopic 
conductivity we should compare 
$R_{N}\sim R $ with the characteristic distance $L$ 
at which VRH conductivity
self averages, or in other words, becomes uniform. This is 
the characteristic period of the critical subnetwork (percolation
infinite cluster) of the Miller-Abrahams resistor network,
which carries all the current\cite{SE}. 
It is known\cite{SE} to be of the order 
of $L = (a/2)(T_0/T)^{(1+\nu_3)/2} 
\simeq a(T_0/T)^{0.95}$. (Here we used the fact that exponent of the 
correlation length of the percolation 
theory, $\nu_3 = 0.9 \simeq 1$\cite{SE}.) The length $L$ 
is of course much larger than 
the characteristic length of a hop $r_h= (a/2)(T_0/T)^{1/2}$.
It is interesting that $L$ is only somewhat smaller 
than the linear screening radius $R_s$.

Below we consider 
two different cases, $R > L, R_s$ and $R < L, R_s $.
In the first case, $R_N > R_s$  and, therefore, screening is linear (
$R_N$ plays no role).
The hole is not bound to a particular donor but 
freely  travels in the spherical layer 
(atmosphere) between $R$ and $R + R_s$. 
In this case, one
can use macroscopic approach for 
the role of traps, namely one can
say that the concentration of electrons (and the local VRH 
conductivity) slightly in varies in spherical atmospheres around all traps.
It is known that in a slightly
inhomogeneous media in the first approximation
the sample conductivity is equal to the average value of the 
local conductivity. 
Therefore, the change of the sample 
conductivity reflects fluctuations of
the total number of conducting electrons and
Eq.~(\ref{nI2}) is valid.

However, the case $R > L, R_s$ can be realized only at
very low frequencies $\nu$ and we have to consider
 the second case, $R < L, R_s $.
In this case,
the simple macroscopic approach based on
 fluctuations of concentration does not work,
because traps perturb the critical network locally,
at the distances much smaller than the 
scale of self-averaging of the VRH conductivity.
We show below that for $R < L, R_s $
$1/f$ noise exists even for the perfect symmetry 
of the density of states with respect of the Fermi level.

\subsection{Effect of the trap potentials}

In the case $R < L, R_s $  
nonlinear screening is important because 
$R_N < R_s$. Therefore, the hole is bound to the
trapped electron and is localized on
a particular donor. In other words, electron 
travels forth and back between
this donor and the trap creating the dipole potential 
$\phi(R, t)$, which modulates conductances $G_{ij}$ 
of neighboring resistors 
of the Miller-Abrahams resistor network\cite{SE}.
In order to calculate the effect of this potential
let us devide the sample in cubic blocks 
with linear size $L$ each and consider for simplicity such
frequencies $\nu$, when $R < L $, but is still 
comparable to $L$, say $R\sim L/5$.
Only very small fraction of these cubes of the order of  
$(dW/d\nu)\nu$ has one trap inside. 
When an electron is captured or 
released, the dipole potential $\phi(R, t)$
changes energy of critical 
resistors which derermine conductance 
of two halfs of the block on both sides of the trap
by energy $e\phi(R, t)= e^{2}/\kappa R \sim k_{B}T$.
Positive and negative potentials
of the dipole applied to 
two different halfs of the cube, 
produce different effects because 
two halfs have different
random conductivities (conductivity self averages only at 
a distance larger than $L$). 
As a result the conductance of 
such cube changes by 100\%.
Thus, due to the trapping and detrapping
of an electron the  
conductance of the cube with 
the trap inside fluctuates roughly 
speaking by factor two. 
Macrosopic conductivity then fluctuates according to 
Eqs.~(\ref{Hooge}) and (\ref{alpha}). 

\section{Comparison with other theoretical models and experiment}

Let us start from two
theoretical papers, which aim at the theory of $1/f$
noise in the VRH conduction. Kozub\cite{Kozub} 
suggested seemingly different mechanism
related to {\it pairs} of donors slowly exchanging 
electron with each other and modulating transport 
on current paths by their potential.
To produce effect at a very small frequency,
two donors should be far from each other. 
But they also should not be "shortened"
by a chain of other donors providing 
faster exchange between them. In other words,
they should be isolated from the conducting media. One could think that
each of two donor should have a pore similar to ones 
discussed in our paper around it. In this case, for a given frequency 
a pore for a "modulator"
pair would have much smaller probability 
than spherical traps we study and "modulators" would 
produce a negligible 
noise on the background of the noise created by traps.

Actually, a pair is not "shortened" even if only one 
of its donors has a spherical pore around it or, in other
words, this donor is exactly a trap we study above.
The second donor is the donor of the $\delta$-band closest to the pore.
So one can say that we presented here a quantitative 
development of the Kozub's qualitative idea. 
Indeed, the language of 
the second half of Sec. IV is very close to 
that of Ref. \onlinecite{Kozub}.  

In another paper, Shtengel and Yu\cite{Yu}
apparently deal with the same mechanism as discussed in our paper. 
They did not study analytical asymptotic dependencies, 
but evaluated expression for $I^{2}_{\omega}/I^{2}$
numerically and obtained results dramatically different 
from ours. They found stronger than $1/f$
growth of $I^{2}_{\omega}/I^{2}$ 
at small frequencies and a weak growth 
of the noise with temperature.
I have no explanation for their numerical results.

Let us now compare our predictions with 
experiments on crystalline semiconductors 
concentrating mainly on the temperature dependence.
In the range of VRH exponential growth of $1/f$ noise with decreasing 
temperature following from Eq.~(\ref{alpha}) 
agrees qualitatively
with $1/f$ noise measurements for implantation
doped silicon samples\cite{McCammon}, but disagrees with results on 
bulk silicon crystals\cite{Lee}.
On the other hand, in agreement with Eq.~(\ref{alpha1}), 
$1/f$ noise is almost temperature
 independent for NNH in germanium\cite{Shlimak}.
$1/f$ noise in the wide range of temperatures including
the transition between VRH and NNH was also studied in two-dimensional
n-GaAs channels\cite{Savchenko}. In agreement with our theory
it was observed that $I^{2}_{\omega}/I^{2}$ sharply decreases
with increasing temperature in VRH range, while in
NNH range the temperature dependence becomes weak.

Frequency dependencies are more problematic for the theory,
because deviations in the direction of smaller power or 
saturation at small frequencies 
were not seen in experimental works.
This can be explained for samples with very large resistances.
When $(T_0/T)^{1/2}$ is very large the frequency of saturation
can become too small to be measured. Indeed, 
at $\nu_0 = 10^{12}$Hz and $(T_0/T)^{1/2} = 20$, 
we find that $\nu_c = 2.5~10^3$Hz and $\nu_{min}=2.5~10^{-4}$Hz.
On the other hand, the absence of a
tendency to saturation in moderately
resistive samples with $(T_0/T)^{1/2} \sim 10$
may signal that the actual
mechanism of $1/f$ noise is different 
from the one considered here. Indeed, there are many 
different pseudoground states separated by 
large potential barriers from each other
in the Coulomb glass formed by a 
frozen distribution of interacting 
electrons on random donors\cite{SE,Kogan}.
Different pseudoground states (valleys) 
can have different conductivities
and random walk of the system through a sequence of
these states may lead to $1/f$ noise\cite{Kogan}. 
Logically it is possible that
this noise can be larger than the one 
we estimated, but nobody was able to
evaluate it. 

Until now we have dealt with the hopping
transport, where both roles of conducting media and traps are 
played by localized states
of impurities. There are also systems 
where current is carried by free electrons,
while localized states play the role of traps only.
The best example of such a system is a MOSFET, where
a free two-dimensional gas exchanges electrons
with localized states (traps) of the oxide.
McWorter's theory\cite{McW} assumes that 
density of states of traps is so small that
any localized state of the oxide, which is 
close to the interface and close in energy to 
the Fermi level
plays the role of the trap.
These traps are assumed to be uniformly distributed
with respect to the distance from the interface and, thus,
lead to $1/f$ noise.

McWorter's theory\cite{McW} neglects the possibility that
near a trap, which exchanges electrons with the conducting channel,
other localized states can be located. They  
can provide alternative hopping paths with
exponentially smaller exchange time
and "shunt" the trap.
However, for small enough frequencies and a
large enough density of localized states in oxide  
these paths always exist. 
Then most effective traps consist of a
localized state surrounded by a spherical pore, which 
touches the interface of the channel. 
As in the case of VRH, 
all localized states within 
energy band $\Delta$ should be excluded from the pore, 
while localized states with 
higher energies can stay there.
Therefore, one can say that the
pore is arranged in the four-dimensional
coordinate-energy space.

The theory of this paper can be directly used to 
calculate the probability of such 
traps and therefore the frequency dependence of 
spectral density of noise. We assume that in 
the gap of the oxide the bare 
density of states, $g$ is constant. 
The Coulomb interaction forms the Coulomb gap 
of the width $E_{CG} \sim e^{3}g^{1/2}/\kappa^{3/2}$
in the vicinity of the Fermi level. One has to compare
the energy "width" of the trap $\Delta$ and 
$E_{CG}$. If $\Delta < E_{CG} $ one can use Eq.~(\ref{alpha}) 
On the other hand, when $\Delta \geq E_{CG}$ Eq.~(\ref{alpha}) 
crosses over to Eq.~(\ref{alpha2}). 

In conclusion, this paper explores the role 
of isolated donors as traps 
which modulate the concentration of conducting electrons
and create noise of the hopping conductivity. 
It is shown that in the wide range of frequencies 
traps lead to approximately $1/f$ 
behavior of noise, with the
crossover to saturation at extremely small frequencies.
In the nearest neighbor hopping range, in spite of the 
activation temperature 
dependency of the conductivity, the relative intensity of $1/f$ noise
is almost temperature independent
(see Eqs.~(\ref{Hooge}) and (\ref{alpha1})).
In the variable range hopping regime 
the intensity of $1/f$ noise 
grows exponentially with the 
decreasing temperature according to 
Eqs.~(\ref{Hooge}) and (\ref{alpha}). 

I am grateful to Yu. M. Galperin, 
V. I. Kozub and M. Weissman 
for the critical comments 
on original version of this paper, 
which led to understanding of
the role of the trap potential
mechanism of the noise (Sec. IV). 
I appreciate useful discussions
with A. L. Efros, H-W. Jiang, Sh. M. Kogan,
D. McCammon, Z. Ovadyahu, A. K. Savchenko
and I. S. Shlimak.

This work was supported by the NSF grant DMR-9985785.

\end{multicols}
\end{document}